\documentclass[11pt]{article}

\usepackage{amsmath}
\usepackage{amsthm}
\usepackage{verbatim}
\usepackage{xspace}
\usepackage[breaklinks=true,
            colorlinks = true,
            linkcolor = blue,
            urlcolor  = blue,
            citecolor = blue,
            anchorcolor = blue]{hyperref}

\newcommand{\mailto}[1]{\href{mailto:#1}{\nolinkurl{#1}}}

\usepackage{tikz}
\usetikzlibrary{matrix}
\usepackage{caption}
\usepackage{enumitem}

\title{Retrofitting a two-way peg between blockchains}

\author{
Jason Teutsch\\
\emph{Truebit}\\
\mailto{jt@truebit.io}
\and
Michael Straka\\ 
\emph{Stanford University}\\
\mailto{mstraka2@stanford.edu}
\and
Dan Boneh\\
\emph{Stanford University}\\
\mailto{dabo@cs.stanford.edu}
}

\date{October 15, 2018\footnote{This version, updated in 2019, includes minor changes and corrections.}}

\newcommand{\DOGE}{{\sf DOGE}\xspace}
\newcommand{\WOW}{{\sf WOW}\xspace}
\newcommand{\ETH}{{\sf ETH}\xspace}

\theoremstyle{plain}
\newtheorem{invariant}{Invariant}
\newtheorem*{requirements}{Requirements}
\theoremstyle{definition}
\newtheorem*{defn}{Definition}
\theoremstyle{remark}

\begin{document}

\maketitle

\begin{abstract}
In December 2015, a bounty emerged to establish both reliable communication and secure transfer of value between the Dogecoin and Ethereum blockchains.  This prized ``Dogethereum bridge'' would allow parties to ``lock'' a \DOGE coin on Dogecoin and in exchange receive a newly minted \WOW token in Ethereum.  Any subsequent owner of the \WOW token could burn it and, in exchange, earn the right to ``unlock'' a \DOGE on Dogecoin.

We describe an efficient, trustless, and retrofitting Dogethereum construction which requires no fork but rather employs economic collateral to achieve a ``lock'' operation in Dogecoin.  The protocol relies on bulletproofs, Truebit, and parametrized tokens to efficiently and trustlessly relay events from the ``true'' Dogecoin blockchain into Ethereum.  The present construction not only enables cross-platform exchange but also allows Ethereum smart contracts to trustlessly access Dogecoin.  A similar technique adds Ethereum-based smart contracts to Bitcoin and Bitcoin data to Ethereum smart contracts.
\end{abstract}

\newpage

\tableofcontents

\section{The bridge over time} \label{sec:bridgeovertime}

\begin{quote}
\textit{
``It appears that we desire a world in which interoperable altchains can be easily created and used, but without unnecessarily fragmenting markets and development.''}~\cite{blockstream}

--Blockstream, 2014
\end{quote}

Since the early days of Nakamoto consensus~\cite{Nak08}, the ideal of trustless agreement across blockchains has captured imaginations and inspired innovation.  The intrigue of joint consensus, however, does not diminish the marvel of universal concurrence over a single blockchain.  The Ethereum community, despite its decentralized structure and multi-billion dollar valuation, has largely agreed over time on who owns each ether coin (\ETH)~\cite{ethereum}.  In contrast, the blockchain community to date has not leveraged Ethereum's reliable consensus for bookkeeping ownership of dogecoins (\DOGE), a set of coins which traditionally change hands via an independent network called Dogecoin~\cite{dogecoin}.

The earliest designs of a \emph{two-way peg}, or systematic transfer of assets back-and-forth between consensus-disjoint blockchains, developed independently from smart contracts, however these two concepts were pioneered concurrently within the context of cryptocurrencies~\cite{ethwhite,blockstream}.  Ethereum's smart contract culture, which catalyzed a new class of two-way peg constructions, has not only influenced but institutionalized thinking on this topic.  Indeed, at the end of 2015, a substantial \ETH bounty emerged to build a bridge between Dogecoin and Ethereum~\cite{bountycontract,bountyannouncement}.  This bridge became known as \emph{Dogethereum}~\cite{dogetherpost}.

Let us muse upon the manifest amalgamation of Dogecoin and Ethereum.  While cultural synergies between these two outgoing, viral communities may have brought their constituents together, one can view the Dogethereum phenomenon entirely as a technical challenge peculiar to this combination of networks.  Dogecoin, which began as a joke at the hands of of Billy Markus and Jackson Palmer, quickly turned into a cultural meme featuring an unnamed shiba inu~\cite{dogecoin,dogeontrial}, an idiosyncratic creole language~\cite{creole}, a \DOGE tip bot~\cite{dogetip}, and the moon \cite{Chohan_2017,dogecoinvideo}.  From its outset, Dogecoin was no stranger to contests or fundraisers.  Dogecoin organizers received dozens of submissions for the Dogecoin's incentivized video contest~\cite{dogecoincontest} including its distinguished winner~\cite{dogecoinvideo}, and backed Jamaica's Olympic bobsled team in 2014~\cite{bobsled}.

On the technical front, Dogecoin began as a fork of Luckycoin, which in turn was a fork of Litecoin~\cite{Chohan_2017}.  Like Litecoin, Dogecoin uses a memory-hard proof-of-work based on scrypt~\cite{scrypthard} which dissuades participation through Bitcoin mining hardware due to that hardware's inefficiency.  By September 2014 the network had begun merge mining with Litecoin \cite{mergemine}, meaning that miners running Dogecoin's mining client could simultaneously mine Litecoin blocks.  Over the following year, Dogecoin's core developers became scarce~\cite{dogecoingithubcommits}.  Without technical leadership, Dogecoin's mining protocol essentially froze despite abundant and dedicated support of Dogecoin enthusiasts.

Dogecoin's memory-hard scrypt proof-of-work naturally drew the attention of Ethereum's developer community, whose increasing demands on computing resources pushed the underlying network to its limits.  Ethereum's \emph{gas limit} quantifies the amount of space and computation that the network can perform per block and places the task of checking a Dogecoin proof-of-work well outside the reach of smart contracts.  Buterin calculated that a na\"{i}ve implementation for verifying a single Dogecoin proof-of-work would require roughly 370 million gas distributed over the course of 118 transactions~\cite{But15}, and Guindzberg's April 2018 estimate placed the gas cost of simply relaying all Dogecoin headers to Ethereum at 10,000 USD per day~\cite{superblocks}.  As increases to Ethereum's gas limit pose security risks to the network's underlying consensus~\cite{LTKS15}, checking Dogecoin's proof-of-work embodies a fundamental challenge for Ethereum.

On the surface, then, Dogethereum seems to call for a trifecta of innovations.
\begin{enumerate}
\item \textit{Consensus.} Formally, how does Ethereum receive a message that \DOGE has been transferred under its control and in accordance with Dogecoin's consensus?  Moreover, how does Dogecoin know when the \DOGE is transferred back?  Both Dogecoin and Ethereum's incentivized mining protocols reinforce myopic views of their respective blockchains.

\item\textit{Computation.} As a minimal prerequisite, Ethereum must be able to authenticate messages describing Dogecoin events.  Dogecoin itself relies on scrypt proof-of-work for this purpose, which means that Dogethereum transfers the burden of scrypt verification onto Ethereum.  Teutsch and Reitwie{\ss}ner introduced Truebit~\cite{truebit,TR17} explicitly to address this problem.

\item \textit{Politics.} How can Dogethereum syntactically express and securely enforce ``locked'' \DOGE transfer given Dogecoin's restrictive, Bitcoin-like scripting language?  Adding a new opcode to Dogecoin requires cooperation on the part of Dogecoin miners, Dogecoin's core developers, and possibly even Litecoin miners who incidentally mine Dogecoin as well.  Indeed, any change in consensus procedure requires a so-called ``fork'' wherein miners each voluntarily update their local clients.
\end{enumerate}

\begin{figure}
\begin{center}
\includegraphics[height=170pt]{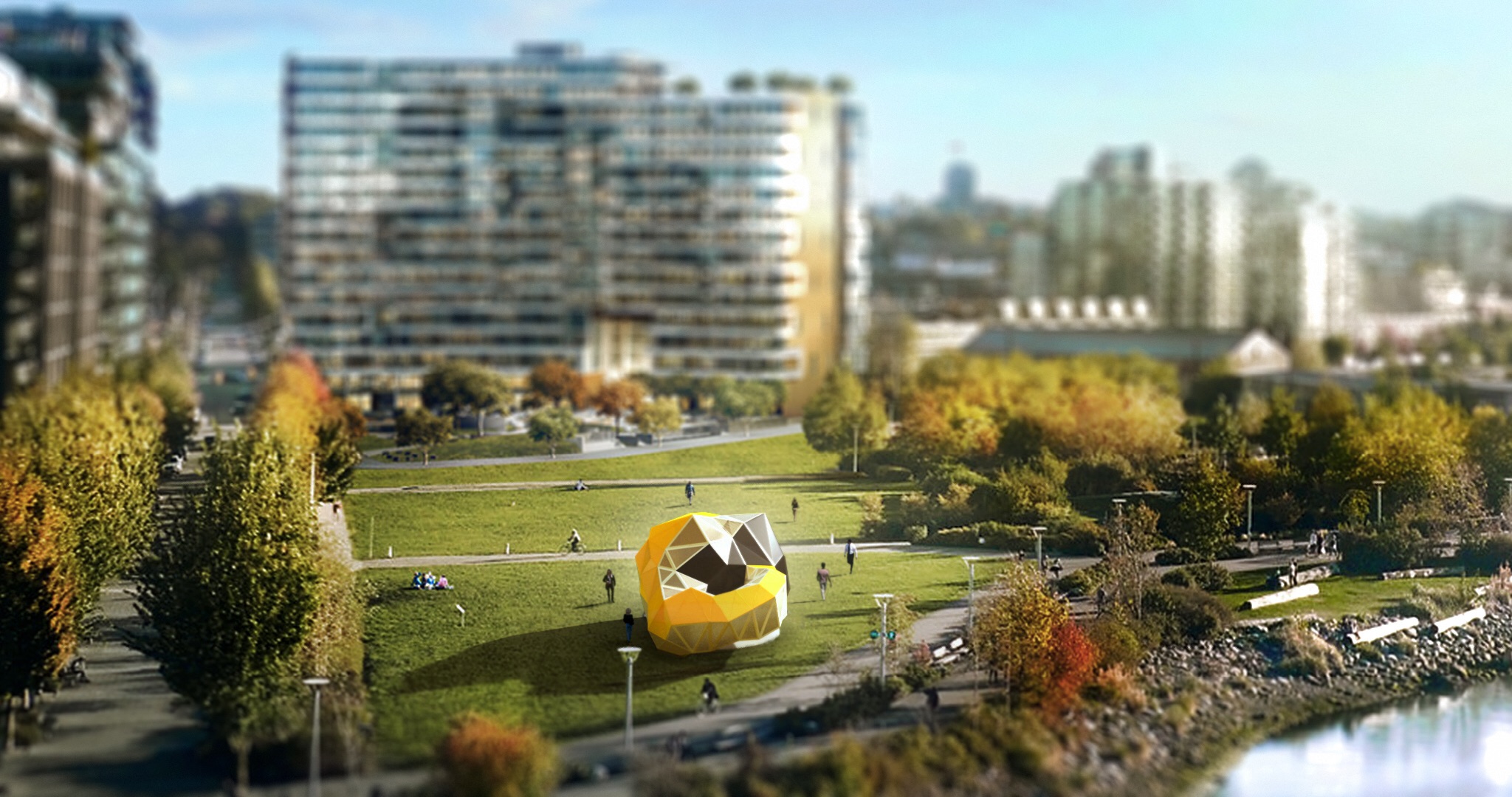}
\end{center}
\caption{Artist rendering of Dogethereum~\cite{artproject}.}
\label{fig:vancouver}
 \end{figure}

By 2017, the stars had aligned on Items~2 and~3 above.  In order to gain the attention of these parties, Truebit introduced the \#ArtProject~\cite{artproject} to create a physical manifestation of Dogethereum in the form of a 40-foot Klein bottle teaming with shiba inus.  With the Truebit protocol implementation \cite{truebitrepo} already underway and a massive Dogethereum bounty~\cite{bountycontract,bountyannouncement} which could potentially finance the \#ArtProject, Dogethereum seemed within reach.  As a proof-of-concept, Truebit's development team munged Chrsitian Reitwie{\ss}ner's scrypt verifier, a hand-crafted tool for playing Truebit-style verification games~\cite{TR17} for scrypt, into a more comprehensive \emph{Truebit Lite} client~\cite{scrypt-interactive} based on altruistic verifiers.  In February 2018, Truebit publicly demonstrated~\cite{dogelivedemo} Truebit Lite alongside Coinfabrik and Oscar Guindzberg's relay prototype, won part of the Dogethereum bounty for their efforts, and committed its entire share of the bounty to the \#ArtProject~\cite{artproject}.  Contrary to press statements at that time, no \DOGE were actually transferred between Dogecoin and Ethereum during this initial demo.

\begin{figure}
\boxed{
\begin{tabular}{ll}
2014:& Dogecoin crowdfunds \$30,000 for Jamaica's Olympic bobsled \\ & team~\cite{bobsled}.\\[1ex]
2015:& Ethereum is born.  Dogethereum challenge inspires conception\\& of Truebit~\cite{truebit,TR17}.\\[1ex]
2016:& Christian Reitwie{\ss}ner commences scrypt verifier~\cite{scrypt-interactive}. \\ & Dogethereum bridge bounty~\cite{bountyannouncement} peaks at 6492 ETH~\cite{bountycontract}.  \\[1ex]
2017:& \href{https://www.reddit.com/r/dogecoin/} {\textsf{/r/dogecoin}} \& \href{https://www.reddit.com/r/ethereum/}{\textsf{/r/ethereum}} merge for April Fool's Day \cite{aprilfools} \\ &following ``Dogether'' announcement~\cite{dogetherpost}.\\[1ex]
2018:& Jessica Angel designs physical manifestation of bridge~\cite{artproject}.  \\ & scrypt verifier is wrapped into Truebit Lite client~\cite{scrypt-interactive}.\\[1ex]
\end{tabular}
}
\caption{Laconic timeline of Dogethereum.}
\end{figure}

\paragraph{Our contributions.}
We describe a ``bulletproof'' two-way peg between Dogecoin and Ethereum with a cryptoeconomic mechanism (Section~\ref{sec:nutshell}) that circumvents the need for either a Dogecoin or Ethereum fork.  Our construction avoids dependencies on politics and distinguished nodes and instead assumes only rational actors (Section~\ref{sec:coc}) atop a simple relay (Section~\ref{sec:bridgeoperation}).  We argue that the construction (Section~\ref{sec:modules}) is sufficiently secure~(Section~\ref{sec:illustrations}) and efficient~(Section~\ref{sec:bulletproofs}) to support practical use.

\subsection{In a nutshell} \label{sec:nutshell}
 
Let us explore the basic operations and applications of our present two-way peg.  As Dogecoin and Ethereum maintain independent blockchains, each of whose public ledgers respectively track ownership of native \DOGE and \ETH coins, we must first establish what it means to migrate \DOGE coins onto Ethereum.  \DOGE transferred onto the Ethereum blockchain takes the form of a \WOW \emph{token}, a coin simulated via smart contract-based ledger named in the spirit of Dogecoin's creole.  \WOW tokens do not follow Ethereum's ERC-20 token standard~\cite{erc20} but rather extend the functionality of this form (see Section~\ref{sec:ingredients} for further details).

We shall introduce the system's agents in Section~\ref{sec:coc} and interact with them more formally in Section~\ref{sec:modules}, but for now we shall simply rely on suggestive names as indicators of roles.  The system does not attempt to manage identities, and as such, a single entity may control multiple agents with distinct names.  In particular, the Hodler may be distinct, or even unknown to the Crosser in the sequence of events below.

The present system introduces neither a ``lock'' opcode nor a preset Dogecoin address for Crossers who wish to send their \DOGE to Ethereum, but rather allows any Operator who deposits sufficient \ETH collateral into a Bridge Contract to facilitate a transfer in exchange for a fee.  A rogue Operator who moves ``locked'' \DOGE forfeits his \ETH collateral.  We refer to this cryptoeconomic incentive as a \emph{collateralized peg} and briefly trace the lifecycle of a \DOGE through this construction.  
\begin{enumerate}
\item An Operator deposits \ETH into a Bridge [smart] Contract in Ethereum.

\item Any Crosser can then ``lock'' \DOGE by sending it into the Operator's Dogecoin address.

\item  A Relayer transmits this ``lock'' transaction to the bridge contract which in turn mints \WOW tokens, in Ethereum, to the Crosser above.

\item When a Hodler \emph{burns} \WOW, provably and irretrievably destroying it,
\begin{enumerate}
\item the Hodler receives ``locked'' \DOGE from an Operator, and
\item the Bridge Contract refunds the Operator's \ETH deposit.
\end{enumerate}
\end{enumerate}
In case the designated Operator fails to ``unlock'' the \DOGE in Step~4, the Hodler collects the Operator's \ETH collateral from the Bridge Contract.  The idea for a collateralized peg was, to the best knowledge of this author, first proposed in a conversation with Sunny Aggarwal and Jim Posen in mid-December, 2017 and shortly thereafter documented in an unpublished manuscript by Sina Habibian \cite{sina}.

\paragraph{Utility.}
Unlike two-way peg constructions which introduce new opcodes, the collateralized peg has the advantage of being retrofitting, making it a relatively frictionless way to bring communities together.  Let us first consider the two-way peg's underlying relay (Section~\ref{sec:bridgeoperation}) in isolation.  
Simply conveying data across blockchains solves a special case of the so-called data availability problem \cite{decentralizedoracles}.  Indeed the relay permits Ethereum smart contracts to seamlessly access data stored on Dogecoin's blockchain.  Thus one can view the collateralized peg itself, which permits cross-chain transfer of value, as a particularly interesting application of the underlying relay.

Dogethereum permits \WOW holders to use their tokens in Ethereum smart contracts, thereby adding a kind of smart contract functionality to Dogecoin.  An ecosystem of two-way pegs gives rise to some interesting applications.  A further two-way peg between Bitcoin and Ethereum, for example, would permit use of tokenized Bitcoins in smart contracts and hence one could trustlessly execute exchanges and interactions between ``tokenized bitcoin'' and \WOW.  With two-way pegs between other systems supporting smart contracts, one no longer has to commit to operating over a single, ``centralized'' decentralized platform like Ethereum, but can catapult tokens onto other blockchains.

Tokens initialized on Ethereum could move elsewhere across a two-way peg oriented in the reverse direction, thereby allowing new blockchains to benefit from existing network effects and communities.  Token holders on the new blockchain may freely transfer tokens acquired on the new blockchain back to Ethereum, even before a market for the blockchain's native coin(s) has been established.  In case a token crosses from Blockchain~$A$ to Blockchain~$B$ to Blockchain~$C$, and then completes a circuit back to~$A$, each individual smart contract on Blockchain~$A$ can decide how to handle the returning version of the tokens.

\paragraph{Open problems.}
The security of the relay in the present construction relies upon specific features of Nakamoto consensus.  An interesting direction for further work would be to design a two-way peg targeting another kind of consensus.  For example one could view a two-way peg between Ethereum and a high-throughput, sharding-based protocol, like Zilliqa~\cite{zilliqa-web,zilliqa-white}, as an alternative to using state channels for micropayments~\cite{funfair, raiden, counterfactual-white, pisa}, although the limited expressibility of Zilliqa's smart contract language complicates matters.  In a similar vein, one might like to securely import data from a high-volume, immutable database like BigChainDB~\cite{bigchain-white} for use in smart contracts.  Building a relay from a non-Nakamoto consensus based system, like Algorand~\cite{algorand}, Dfinity~\cite{dfinity}, Tezos~\cite{Tezos}, Thundercore~\cite{thundercore}, PHANTOM~\cite{Sompolinsky2018aa}, or a collateralized peg from a privacy-enhanced blockchain like Zcash~\cite{zcash-web} or Monero~\cite{monero} offer additional technical challenges.   Moelius's Ethereum Input Bus~\cite{eib-convo,eib-github}, when adapted to support ledgers and logs which guarantee data availability, may offer a useful data verification tool for securely extracting external bits of information in this context.

\subsection{Ingredients} \label{sec:ingredients}

We describe the (not-so) secret sauces which combine to yield Dogethereum.  First, we introduce the notion of \emph{parametrized token} which extends the ERC-20~\cite{erc20} token standard.  Parametrized tokens shape the incentives for Dogethereum's collateralized peg, while Truebit and bulletproofs (see below) constitute the core of the relay which conveys Dogecoin events onto Ethereum.

\paragraph{Parametrized tokens.}

The utility of a two-way peg between Dogecoin and Ethereum could increase the value of \DOGE, which in turn could render an Operator's \ETH collateral insufficient relative to its ``locked'' \DOGE.  In order to make the bridge robust against changes in exchange rates, we create a separate \WOW token for each possible exchange rate, each indexed by a  positive, rational number, i.e.
\[
\WOW[1], \WOW[2], \WOW[1/2], \WOW[4], \WOW[1/4], \dotsc.
\]
A \emph{parametrized token}, then, is an ERC-20 token with quantitative parameter(s) attached to it.  In the case of \WOW, the parameter reflects the maximum ``safe'' exchange rate with sufficient \ETH backing.  Parametrized tokens differ from the emerging non-fungible ERC-721 standard~\cite{erc721} in that parametrized tokens with identical parameters are all interchangeable whereas ERC-721 assumes that each token is distinct.  When clear from context or when the discussion does not involve exchange rates, we shall informally refer to $\WOW[y]$ tokens as ``\WOW'' without specifying the parameter~$y$ explicitly.  

Inevitably, when the value of a currency decreases relative to another, some party must bear this change.  A Hodler holding \WOW is free to exchange it for \DOGE at any moment and, hence justly absorbs \WOW's price upside gain as well as its downside. Operators, on the other hand, accept some \ETH illiquidity while keeping custody of ``locked'' \DOGE in exchange for obtaining a front-loaded crossing fee.

We shall observe, via a series of invariants (Section~\ref{sec:reserve}), that the two-way peg maintains balance, and hence liquidity, between the quantity of ``locked'' \DOGE and the $\WOW[y]$ in circulation, so long as the ``true'' value of 1~\ETH remains above~$y$.  In case the \ETH exchange rate drops below this threshold, the protocol does not guarantee such balance; Hodlers must exchange their $\WOW[y]$ for \DOGE before this happens.    On the flip side, Operators can, at any moment, keep their ``locked'' \DOGE at the cost of forfeiting \ETH collateral.  We shall introduce a further pair of token parameters for deposits in Section~\ref{sec:deposits}.  Zamyatin, Harz, Lind, Panayiotou, Gervais and Knottenbelt concurrently developed a collateralized, two-way peg construction which mitigates exchange rate fluctuations without parameterized tokens and instead relies on a designated price oracle~\cite{ZHLPK18}.
\paragraph{Bulletproofs.}

Dogethereum must maintain an accurate record of ``locks'' that have occurred on the ``true'' Dogecoin blockchain.  The Bridge Contract democratically permits anyone to update this record, however each update must pass by unanimous consensus.  In case of a dispute, the active Relayer who posted the update must provide a proof of its correctness.  We use bulletproofs to succinctly prove validity of Dogecoin proof-of-work, a non-trivial task~\cite{superblocks,But15}, and thereby achieve a space-efficient, non-interactive relay enforcement on Ethereum.  We further reduce space and time requirements by using batch verification techniques to quickly verify entire chains.  A \emph{bulletproof}~\cite{bulletproofs, BBBPWM17} is an efficient certificate that a committed arithmetic circuit is satisfiable.  The size of a bulletproof, as noted in the precursor work~\cite{BCCGP16}, is only logarithmic in the circuit size~$n$, while proof generation and verification run times are asymptotically linear in~$n$.    Bulletproof implementations exist and are ready for use~\cite{libsecp256k1, rust-bulletproofs}.

Related proof systems for verifiable computing include SNARKS~\cite{libsnark,BCGTV13}  and STARKS~\cite{starkware,starks}.  Compared to SNARKS, which require a trusted setup~\cite{Wil16,zcashceremony}, and STARKS, whose proof size starts at 200 kB and whose prover memory requirements are especially high for memory-hard scrypt~\cite{ BBBPWM17}, bulletproofs are relatively efficient and avoid trusted setup~\cite{Poe18}.  While each of these proof systems showcase privacy features, we remark that Dogethereum does not involve any use of zero-knowledge.

\paragraph{Truebit.}

\emph{Truebit} \cite{truebit, TR17} is an on-chain, pay-per-use, computation oracle which trustlessly circumvents Ethereum's gas limit.  The present two-way peg construction uses Truebit to verify bulletproofs.  We shall treat Truebit as a black box.

\subsection{Related work} \label{sec:relatedwork}

Diverse ideas on two-way pegs have surfaced over the years with various advantages and tradeoffs.  As we view this work in the context of existing literature, let us keep in mind that our primary aim is to fully specify a retrofitting two-way peg composed of rational and Byzantine agents, as opposed to reputation-based ones.

An \emph{atomic swap} is a cross-chain coin exchange protocol between two parties where either both successfully receive each other's coins or neither of them do.  Atomic swaps go hand-in-hand with Dogethereum because they facilitate access to \WOW and hence add liquidity to the system.  Indeed they offer an alternative way to trustlessly exchange \DOGE for \WOW without the complications of participating in Dogethereum itself.   While traditional atomic swaps require each participant to explicitly find a trading partner prior to executing the swap~\cite{tiernolan}, recent constructions may add liquidity through reserves~\cite{commonwealthcrypto}.  As we shall see, Dogethereum requires no pre-designation of a trading partner.  Unlike two-way pegs, however atomic swaps do not transfer data or consensus across blockchains.

One way to achieve a two-way peg is for one chain to adapt a consensus mechanism which treats the other chain's consensus as authoritative.  If Dogethereum were to follow this model, Dogecoin might introduce a hard fork with new opcodes for ``lock'' and ``unlock'' and Dogecoin miners would agree to monitor Ethereum for \WOW that gets burned.  See \cite{rsksidechainblog,rsktechpaper} and \cite{blockstream} for more detailed examples of such \emph{sidechain} constructions which rely on \emph{simplified payment verification} (a.k.a\ SPV proofs)~\cite{bitcoin-white}.  Kiayias, Miller, and Zindros recently introduced a concise form of SPV proofs called ``Non-interactive proofs of proof-of-work'' (NIPoPoW) \cite{nipopows,KMZ18} which leverage the expected existence of \emph{superblocks}~\cite{highvaluehash}, blocks whose proofs-of-work satisfy higher hash difficulty.  We remark that using NIPoPoWs in a Dogethereum construction would require a Dogecoin fork and entail further design considerations.

One can realize a forkless two-way peg by appealing to external governance.  A  \emph{federated peg}, for example, appoints a set of addresses with the authority to lock, unlock, and perhaps  even mint currency.  Users of a federated peg trust the parties holding these addresses to follow the designated protocol honestly.  The assumption that say, $k$ out of $n$, of the authorities will not collude may, in practice, suffice based on geographic, political, and reputational considerations.  RSK employed this technique, which they call ``multi-sig federation,'' in their 2-way peg with Bitcoin~\cite{rsksidechainblog}.  Their construction also uses a \emph{drivechain} in which miners of one network vote on the locks of the other.  In a drivechain scenario, Ethereum miners might observe the Dogecoin blockchain and indicate in their blocks whether or not each \DOGE is locked, and the managing Ethereum smart contract would mint a \WOW once confirmed through a threshold of votes.  This construction relies on honest participation of miners, and without a formal fork on Ethereum to incentivize correct behavior, miners might become lazy about participating in the drivechain or susceptible to bribes.  Polkadot~\cite{polkaweb,polkapaper} offers an alternative method for connecting blockchains by allowing participant networks to delegate consensus via a Relay Chain.  Presumably a Polkadot-style bridge between Dogecoin and Ethereum would involve third-party consensus from the Relay Chain.  The Cosmos Hub~\cite{cosmoshub} follows a similar approach.

BTC Relay~\cite{btcrelay} was an early experiment by Consensys and Ethereum which incentivized participants to convey Bitcoin block headers to a smart contract, thereby enabling SPV verification of Bitcoin events in Ethereum.  This methodology allows Bitcoin transactions to trigger events in Ethereum but falls short of being a two-way peg as it does not transfer control of value across blockchains.  In contrast to this mono-directional scheme, Peace Relay~\cite{peacerelay} bridges Ethereum and Ethereum Classic in a bi-directional manner.  This construction harnesses the smart contract functionality in both blockchains in order to achieve a two-way peg.  Peace Relay uses the efficient 
Ethash verification technique introduced in Smartpool~\cite{smartpool} to check proof-of-work from each of these blockchains without appealing to Truebit.  BTC Relay and Peace Relay both include working code implementations but do not address the issue of orphaned block submissions (see Section~\ref{sec:bridgeoperation}).  Indeed, Ethereum can only learn whether a given block belongs to the ``true'' Dogecoin blockchain \emph{a posteriori}, as Ethereum miners only reach consensus on Ethereum events.

Bejarano and Guindzberg's ``superblock'' relay construction~\cite{superblocks}, which shares the same name as the NIPoPoW-related concept above but bears no technical resemblance to it, explicitly tackles the issue of orphaned block submissions.  Roughly speaking, their relay simulates Dogecoin's Nakamoto consensus inside of an Ethereum smart contract in order to keep track of  ``lock'' events on Dogecoin.  Rather than relaying each individual Dogecoin block header to Ethereum, Bejarano and Guindzberg batch the submissions into Merkle-hashed ``superblocks'' in order to save gas.  Only in case of a challenge are the Dogecoin block headers represented the superblock revealed on-chain, in which case the relay appeals to Truebit to check each scrypt proof-of-work.  The protocol weeds out orphaned blocks by penalizing challenged blocks which do not make it into the simulated ``main chain.''   As the smart contract doesn't view the Dogecoin block headers in absence of challenges, however, no canonical main chain may exist in case of two incomparable but unchallenged forks (e.g.\ from two extensions with incomparable partitions of time stamps).  In general, the Nakamoto consensus simulator must have a formal means of determining the main chain, as either there is moment in which the Bridge Contract generates \WOW or there isn't.

In September 2018, Guindzberg and CoinFabrik publicly demonstrated \cite{sept5demo} their superblock implementation~\cite{oscar-repo} to the Dogethereum bounty judges as a follow up to their joint presentation with Truebit earlier that year~\cite{dogelivedemo}.  Both demos included Truebit's specialized scrypt verifier and Truebit Lite \cite{scrypt-interactive,sinalive}, a semi-altruistic incentive layer and client.  The present paper aims for a simpler and more robust relay construction with efficient verification, in terms of cost of deposits, bits communicated on-chain, and speed of confirmation.  In contrast to the superblock construction~\cite{superblocks} above, the present relay construction permits Dogecoin extensions to be submitted sporadically as opposed to at hourly intervals, confirms crossings in minutes rather than hours, and minimizes Relayer deposits.

\subsection{Blockchain lingo}

We review some technical notions used in our construction.  A \emph{blockchain} is an immutable, public ledger which grows as \emph{miners} add a linear ordering of \emph{blocks} to it over time.  Dogecoin blocks contain \emph{transactions} which transfer \DOGE coins from one address to another, while Ethereum blocks contain state updates for \emph{smart contracts}~\cite{ethereum,nick-szabo}, or small computer programs which process tokens, coins, and data.  A decentralized process called \emph{Nakamoto consensus}~\cite{bitcoin-white} determines which blocks get added to the blockchain in each epoch lasting approximately  14~seconds in Ethereum~\cite{ethblocktime} or 62~seconds in Dogecoin~\cite{dogeblocktime}.  Each valid block includes a correct solution to a hard cryptographic problem, tied to it predecessor, called a \emph{proof-of-work}.  The metadata for a block, which contains a hashed commitment to its transactions but not a full list of them is called a \emph{header}.   A \emph{fork} is either a modification or disagreement in underlying consensus and can refer either to an event which changes the consensus mining protocol or one which results in incompatible extensions in the underlying blockchain.  The word ``fork'' can also refer to one of the incompatible extensions itself.

The \emph{longest chain rule} refers to the convention in Nakamoto consensus that the valid chain with the greatest cumulative difficulty, typically the longest one, is the ``correct'' one for miners to extend and which eventually becomes permanent on account of its universal acceptance.  A valid block which ends up on a fork rather than the permanent blockchain is called \emph{orphaned}.  The first block in a blockchain is called the \emph{genesis} block, and the \emph{ordinal number} of a block is its distance, as measured in transitive proof-of-works, from the genesis block.  

A \emph{Merkle tree} is a binary tree in which each node is the hash of the concatenation of its children nodes~\cite{smartpool}. In general, the leaves of a Merkle tree will collectively contain some data of interest, and the \emph{root} is a single hash value which acts as a certificate commitment for the leaf values in the following sense. If one knows only the root of a Merkle tree and wants to confirm that some data $x$ sits at one of the leaves, then holder of the original data can provide a path, or \emph{Merkle proof}, from the root to the leaf containing $x$ together with the children of each node traversed in the Merkle tree. Such a path is difficult to fake because one needs to know the children’s hash preimages for each hash in the path, so with high probability the data holder will supply a correct path if and only if $x$ actually sits at one of the leaves.   The \emph{Merkle hash} of a data set is the root of a Merkle tree which contains the data at its leaves, and we describe such a data set as \emph{Merklized}.

We shall refer to the ``true'' Dogecoin blockchain with quotation marks because, at moments where distinct miners broadcast blocks at the same time, one's perception of the longest chain may depend on network viewpoint relative to this pair of blocks.  Similarly, the ``true'' exchange rate between \DOGE and \ETH is always related to a particular agent as no such exact, universal notion exists.  We place quotes around the words ``lock'' and ``unlock'' to remind ourselves that Operator have the private keys necessary to move ``locked'' \DOGE at any moment and, from a cryptographic point-of-view, need not wait for an ``unlock'' event.  Finally, we shall use the words ``Dogecoin'' and ``Ethereum'' to refer to both communities and networks without explicit declaration when the distinction is clear from context.

\section{Casting and behavior profiles} \label{sec:coc} 

We give an overview of the agents in our collateralized peg.  As hinted in Section~\ref{sec:nutshell}, and as we shall explore fully in Section~\ref{sec:modules}, our Dogethereum construction assumes neither distinguished nor honest nodes, although we do assume rationality as described below.  The \emph{Bridge Contract}, which is smart contract in Ethereum, manages the overall function and ensures key properties of the system.  

A single entity may simultaneously fill multiple of the roles below and, in fact, our model predicts some overlap.  Actors need not trust other agents or rely on correctness of systems components other than the Bridge Contract, Dogecoin's consensus protocol, bulletproofs, and Truebit.  In Section~\ref{sec:reserve}, we shall argue that the system's incentives are robust against collusion.  We make the following security assumptions.
\begin{enumerate}
\item \emph{Block withholding attacks} \cite{TJS17},  in which miners find valid blocks but do not broadcast them, do not occur.  As the two-way peg protocol does not involve randomness outside of witnesses for bulletproofs, block withholding is unlikely to bias operations.

\item Attackers have limited computational resources and cannot execute \emph{double spends} or \emph{51\% attacks} \cite{bitcoin-white,TJS17}, that is, they cannot erase or modify transactions recorded on the Dogecoin blockchain.

\item Parties may communicate off-chain, and \emph{Sybil attacks}, which convolve identities, are expressly permitted.

\item Attackers will not waste significant financial resources on denial-of-service attacks on either the Bridge Contract or Ethereum; we ignore potential profitability from such attacks outside the present closed system.  Moreover, we assume attackers cannot censor smart contract transactions.  In general, censoring specific actions is difficult because each smart contract can call other smart contracts.
\end{enumerate}

At a high level, the collateralized peg is a zero-sum game.  Agents have incentive to monitor each other's actions since some agent loses whenever another one cheats.  Without further ado, let's meet the characters: Operators, Crossers, Hodlers, Relayers, and Reporters.  We shall refine these roles further in Section~\ref{sec:modules}.  In the descriptions below, \emph{history} refers to the sequence of Dogecoin blocks known to the Bridge Contract at a given moment.

\subsection{Operators}
An \emph{Operator} facilitates cross-chain consensus hand-offs between \DOGE and \WOW by providing \ETH backing against newly generated \WOW tokens.  The operator stores ``locked'' \DOGE in a Dogecoin address which she fully controls but can lose her \ETH collateral if she moves these \DOGE prematurely, that is, before certain \WOW tokens get provably burned.  The \WOW tokens generated through the ``lock'' process (Section~\ref{sec:locking}) need not be the same as those burned to release the ``lock'' (Section~\ref{sec:unlocking}).  The Operator's initial \ETH collateral must exceed the value of her ``locked'' \DOGE.  We now characterize the Operator's demeanor.

\begin{description}
\item[\textmd{\emph{Reasons for participation.}}]\ 
\begin{enumerate}[label =\arabic*.]
\item Operators receive front-loaded \emph{crossing fees}, paid in \WOW, for facilitating crossing (see Section~\ref{sec:bridgeopening}).

\item Operators have bounded capital risk.  If the value of the ``locked'' \DOGE rises above the value of the Operator's \ETH collateral, the Operator may keep the ``locked'' \DOGE instead of the \ETH collateral.  In case of more favorable exchange rates, the Operator can eventually retrieve back her \ETH collateral on a first-in-first-out basis (Section~\ref{sec:unlocking}).  The Bridge Contract never requests the Operator to add more \ETH than her initial collateral deposit.
\end{enumerate}

\item[\textmd{\emph{Assumptions.}}]\ 
\begin{enumerate}[label=\Alph*.]
\item Operators have access to initial \ETH capital to place as collateral into the Bridge Contract.

\item Operators may not mind holding illiquid \ETH collateral for a while but wouldn't wish to leave their funds in a smart contract indefinitely.  Operators eventually want their \ETH back, plus some interest.  For this reason, the protocol institutes a first-in-first-out queue for handling \ETH collateral (Section~\ref{sec:minting}).  In case an Operator's bridge sits unused due to the exchange rate or otherwise, the Operator can cross it himself, burn the \WOW, and retrieve both his ``locked'' \DOGE and \ETH collateral.

\item Operators control some Dogecoin address and can interact with Ethereum smart contracts.  The Operator must provide a Dogecoin address at which to ``lock'' and ``unlock'' \DOGE, and must also provide \ETH collateral to the Bridge Contract (Section~\ref{sec:bridgeopening}).

\item Operators may monitor the ``true'' exchange rate between \DOGE and \ETH and withdraw ``locked'' \DOGE from the system in accordance with such observations at the cost of forfeiting \ETH collateral.

\item Operators may wish to monitor the Bridge Contract through participation as a Relayer (see Section~\ref{sec:relayers} below) in order to avoid false claims that ``locked'' \DOGE has been moved or was not released in a timely manner.

\item While it may be in a rational Operator's best interest to prefer \ETH or \DOGE at a given moment, Byzantine choices on the part of the Operator do not harm the incentive or decrease net value for other agents (Section~\ref{sec:reserve}).
\end{enumerate}
\end{description}

\subsection{Crossers}

\emph{Crossers} initially hold \DOGE coins and wish to obtain \WOW tokens.  A Crosser may deposit \DOGE into any address controlled by an Operator (Section~\ref{sec:locking}) and, upon confirmation of the deposit via relay (Sections~\ref{sec:relay-listening} and~\ref{sec:relay-verification}), receive \WOW from the Bridge Contract in return (Section~\ref{sec:minting}).  Upon receipt of the \WOW, the Crosser transmogrifies into a Hodler (Section~\ref{sec:hodlers}).

\begin{description}
\item[\textmd{\emph{Reasons for participation.}}]\ 

\begin{enumerate}[label =\arabic*.]
\item Crossers wish to obtain \WOW for use in smart contract operations.

\item Crossers may intend to resell \WOW to others.
\end{enumerate}

\pagebreak

\item[\textmd{\emph{Assumptions.}}]\ 
\begin{enumerate}[label=\Alph*.]

\item Crossers are willing to pay a modest fee in order to exchange \DOGE for \WOW.  Thus, each Crosser initially holds \DOGE and must have an \ETH address at which to receive \WOW.

\item Crossers have \DOGE for crossing but also may wish to have on hand a small amount of \ETH to stake temporarily for additional security (Sections~\ref{sec:simultaneous} and~\ref{sec:locking}).  Crossers who wish to take advantage of this feature must be able to send data to smart contracts.

\item Crossers monitor ``true'' exchange rates in order to ensure that the Operator's \ETH collateral suffices to cover ``locked'' \DOGE.  Crossers can decide whether or not to use a bridge based on the Operator's fees and threshold rate for the \ETH collateral (Section~\ref{sec:bridgeopening}).

\item Crossers may monitor the Bridge Contract, in the short run, in order to ensure that their \DOGE ``locking'' is properly recorded in the Bridge Contract's history, i.e.\ that the Dogecoin ``lock'' transaction is neither omitted nor manipulated so as to pay newly minted \WOW to an attacker's Ethereum address.  In order to defend against such vulnerability, the Crosser may participate as a Relayer (Section~\ref{sec:relayers}).

\item Crossers can also be Operators.  An agent that acts as both an Operator and a Crosser can pass \WOW to herself in order to maintain liquidity against her ``locked'' \DOGE.
\end{enumerate}
\end{description}

\subsection{Hodlers} \label{sec:hodlers}

Any party holding a \WOW token is a \emph{Hodler}.  Hodlers who burn \WOW into the Bridge Contract wish to obtain \DOGE, and the oldest Operator is obligated to provide this \DOGE to the Hodler, lest the Hodler earn the right to collect that Operator's \ETH collateral (Section~\ref{sec:reportingmissing}).  A Hodler can either be a person or a smart contract (see Section~\ref{sec:dex}).

We assume that Hodlers track the ``true'' \ETH to \DOGE exchange rate.  Empirically, we observe many hodlers doing this in the real world!  In particular, Hodlers will recognize when the margin supported by the \ETH collateral becomes insufficient, in which case they must cross their \WOW back to Dogecoin to avoid potential loss of value (Section~\ref{sec:unlocking}).  Hodlers have incentive to act as Relayers and monitor the Bridge Contract's history whenever they cross \WOW into \ETH because a malicious Operator could lie to the Bridge Contract about transmitting \DOGE to the Hodler's Dogecoin address.  In short, Operators absorb exchange rate fluctuations up to the $\WOW[y]$'s threshold exchange rate of 1 \ETH = $y$ \DOGE, while Hodlers bear the effects of market fluctuation whenever the price of \DOGE exceeds this bound.  As Hodlers can exchange \WOW for \DOGE at any moment, participation in the two-way peg incurs minimal liquidity risk for them, whereas Operators bear only bounded risk relative to the exchange rate.

Hodlers have an incentive to track Dogecoin transactions so that the can report missing \DOGE in exchange for rewards in the form of \ETH collateral (Section~\ref{sec:reportingmissing}).

\subsection{Relayers} \label{sec:relayers}

\emph{Relayers} convey ``locks'' and ``unlocks'' in Dogecoin to the Bridge Contract in Ethereum.  Relayers engage in a unanimous consensus protocol (see Sections~\ref{sec:relay-listening} and~\ref{sec:relay-verification}), which means that security scales with increased participation.  Each Relayer posts a deposit (see Section~\ref{sec:deposits}) which affords the right to both submit a Dogecoin extension to the Bridge Contract's history (Section~\ref{sec:relay-listening}) and to challenge the validity of another Relayer's submission (Section~\ref{sec:relay-verification}).  Deposits cover the expense of resolving disputes while acting as a deterrent against denial-of-service.

\begin{description}
\item[\textmd{\emph{Reasons for participation.}}]\ 

\begin{enumerate}[label =\arabic*.]
\item As described above, Operators, Crossers, and Hodlers each have skin in the game at various times and may decide to participate as Relayers in order to avoid being cheated.  Since Relayers only submit Merkle hashes to the Bridge Contract,  individual agents can only recognize an incorrect Merkle hash but can't tell \emph{a priori} whether they are personally affected by its encapsulated attack.  Therefore there exists an incentive for every vulnerable agent to challenge every bogus Merkle hash.  Even if the full contents of the Merkle hash were to be broadcast publicly off-chain, there is always at least one potential victim who has explicit reason to challenge.  Even if Operators and Crossers were to conspire to get the relay stuck (Section~\ref{sec:backtracking}), Hodlers might wish to prevent the system from derailing.

\item Relayers who successfully extend the Bridge Contract's history receive relay tax from each Crosser who sent \DOGE to an Operator in the associated blocks (see Section~\ref{sec:locking}).
\end{enumerate}

\item[\textmd{\emph{Assumptions.}}]\ 
\begin{enumerate}[label=\Alph*.]

\item Relayers each supply a deposit in \ETH which is distinct from any Operator deposit.  This separation ensures that relay activities do not jeopardize the collateralized peg invariants.

\item Relayers can view the ``true'' Dogecoin blockchain, compute Merkle hashes, and send data to smart contracts.  After all, the purpose of the relay is to inform the Bridge Contract about Dogecoin events.

\item Individual Relayers may not always be online, and need not monitor Dogecoin continuously.  A Relayer's deposit is only at stake at the time when he proposes an extension to the history or challenges someone else's.

\item A relay tax (Section~\ref{sec:locking}) helps to avoid inertia and a tragedy of the commons in which lazy Operators, Crossers, and Hodlers each wait for someone else to relay information from Dogecoin to Ethereum.
\end{enumerate}
\end{description}

\subsection{Reporters}

\emph{Reporters} provide Merkle proofs  of ``locks'' and ``unlocks'' stored in the Bridge Contract's history to the Bridge Contract itself in exchange for rewards.  Reporters need not post deposits as the Bridge Contract can itself verify Merkle proofs and simply ignore them if incorrect or redundant.  Direct verification minimizes space for attack surfaces.  There are two sources from which Reporters may collect fees:
\begin{enumerate}
\item Operators may offer \emph{burn reporting bounties}, established at the time of bridge creation, to ensure that the eventual release of \DOGE gets reported to the Bridge Contract (Sections~\ref{sec:bridgeopening} and~\ref{sec:minting}).

\item Crossers may offer \emph{lock reporting bounties} to ensure that the Bridge Contract learns of their \DOGE ``locks'' (Sections~\ref{sec:locking} and~\ref{sec:minting}).
\end{enumerate}

Reporters improve user experience for Operators, and Crossers by automating certain tasks, however they are not otherwise essential to the protocol.  Agents who wish to perform reporting tasks themselves need not offers fees to Reporters.  Although incentives exist for anyone to participate as a Reporter, we especially expect Operators, Crossers, Hodlers, and Dogecoin miners to fill these roles because of the the marginal extra work required beyond the monitoring they already do on Dogecoin.

\section{Relay design philosophy} \label{sec:bridgeoperation}

At the heart of any two-way peg lies a relay routine which transfers data and consensus across blockchains.  Our relay construction conveys ``lock'' and ``unlock'' events from Dogecoin to Ethereum, although a similar method could be used to access any data stored on Dogecoin's blockchain.  We remind the reader here that ``lock'' and ``unlock'' events look no different from any other transaction in the eyes of Dogecoin, however the Bridge Contract can interpret them as such based on \ETH collateral deposits and \WOW burns that it witnesses on the Ethereum side.  Here we shall concern ourselves strictly with the relaying of blocks without attaching any semantic meaning to the information contained in individual transactions.

We wish for the Bridge Contract to unerringly accept a monotonically increasing sequence of extensions which mimics blocks on the ``true'' Dogecoin blockchain.  Throughout this discussion, we shall assume that an adversary wishes to exploit the relay by confirming bogus Dogecoin data onto the Bridge Contract.  Let us first consider an attack and a strawman solution which will inform our ultimate design choices.

\paragraph{\textmd{\it Orphaned block submissions.}}  Suppose that a Relayer were to supply the Bridge Contract with an extension of valid blocks which do not appear on the ``true'' Dogecoin blockchain.  If the Bridge Contract were to accept such an extension, not only might bogus ``locks'' and ``unlocks'' occur that adversely affect various agents, but the relay might become non-extendable to honest Relayers following proof-of-works on the ``true'' Dogecoin blockchain.  A Relayer could introduce such an error either accidentally or opportunistically after mining a valid block that failed to reach confirmation on Dogecoin; the Bridge Contract must reject such orphaned block submissions.

\paragraph{\textmd{\it Sampling.}}
An extension may include many Dogecoin blocks.  Due to the prohibitively high cost of communicating a full sequence of Dogecoin blocks to Ethereum, let alone verifying each of them, one might be tempted to simply allow the Relayer to provide a Merkle hash of the extension and sample a block or two from the extension in order to confirm the validity of its constituents.  Unfortunately, searching for an error in this way is like looking for a needle in a haystack.  An adversary could hide a single invalid proof-of-work almost anywhere in the sequence, hence the probability of chancing upon a single error in a long extension is negligible.  Bounding the adversary's mining power doesn't help because any Relayer could copy the ``true'' Dogecoin blockchain up to any point in the extension and then begin his malicious splice from there.  Blocks to support such an attack could be copied from previous or orphaned Dogecoin blocks, intentionally mined off-line, or even mined in real time.  Once a single incorrect proof-of-work has been injected into the sequence, the subsequent proof-of-works which correctly follow the malicious could hide the attack from a sampling radar.  We emphasize that it does not suffice to simply observe that the adversary's Merkle hash  does not match that of the ``true'' Dogecoin blockchain --- the challenger must somehow prove this to the Bridge Contract.  An Ethereum protocol cannot distinguish between a ``true'' Dogecoin block and an orphaned one simply by checking its validity.

\bigskip

In order to circumvent the problems described in the previous two paragraphs, we shall require the following three properties which suffice to guarantee that an extension belongs to the ``true'' Dogecoin blockchain.
\begin{requirements}
A valid extension submitted to the Bridge must meet the following three properties.
\begin{enumerate}
\item \emph{Maximality.} The extension length must be (very close to) maximal at the time of submission.
\item \emph{Validity.} Every proof-of-work in the extension's sequence of blocks must be valid.
\item \emph{Shallow-fork-free.} The extension is ``confirmed'' via a witnessing sequence of additional, valid proofs-of-works which do not count towards the extension's length.
\end{enumerate}
\end{requirements}
Every block in a correct extension must satisfy Validity, as, according to the longest chain rule, the ``true'' Dogecoin blockchain contains only valid proof-of-works.  Note that by ``extension length'' above we formally mean cumulative difficulty.  Using this definition, any extension of the ``true'' Dogecoin blockchain satisfying Maximality reflects the greatest, cumulative, valid proof-of-work and therefore represents a ``true'' Dogecoin blockchain at some moment.  Moreover, any such extension which is ``confirmed'' via the Shallow-fork free property, permanently belongs to the \emph{``true''} Dogecoin blockchain and avoids including orphaned blocks.  In Sections~\ref{sec:relay-listening} and~\ref{sec:relay-verification}, we shall describe how our two-way peg meets the above requirements (see also the definitions at the beginning of Section~\ref{sec:modules}).

\subsection{Simultaneous crossings} \label{sec:simultaneous}

An inevitable but unfortunate situation arises when two Crossers simultaneously attempt to cross \DOGE via the same Dogecoin address.  The Operator who controls that address may then have more \DOGE than he has backed with \ETH collateral, and a rational Operator would then walk off with the union of the Crossers' \DOGE.  In order to combat this shortcoming, Crossers may optionally register with the Bridge Contract prior to using a bridge in order to uniquely reserve a crossing spot and avoid accidental collision (Section~\ref{sec:locking}).

\subsection{Computational complexity}

The relay must resolve disputes regarding the validity of an extension on-chain.  In such cases, rather than presenting all the block headers on-chain at considerable gas cost from space storage, we rather require the active Relayer in question to supply a laconic bulletproof demonstrating validity of his submission.  This approach also has the nice side effect of being completely non-interactive in the sense that the active Relayer simply submits a single bulletproof to the Bridge Contract, which in turn calls Truebit for binding verification.  The Challenger need only watch this process from the sidelines.  The Relayer's bulletproof roughly takes the following form (Section~\ref{sec:relay-verification}):
\begin{quote}
``I committed a confirmed, valid sequence of Dogecoin blocks extending the latest block known to the Bridge Contract at the time of submission.''
\end{quote}

We remark that the more efficient the arithmetic circuit is for the scrypt proof-of-work bulletproof, the faster and cheaper the proof generation and verification will be (Section~\ref{sec:bulletproofs}).   Due to the relay's threading scheme (Section~\ref{sec:relay-verification}), the relay need not wait for verification resolution before reading in the next extension.  Hence threading substantially lowers the practical speed requirements for proof-of-work verification in Dogethereum.  scrypt bulletproof generation and verification times depend on the total number of blocks in the extension but not the number of transactions per block.

\subsection{System highlights}

Before presenting a formal specification for the two-way peg, let us anticipate and recap some properties of the construction.

\begin{enumerate}
\item \emph{Forkless.} The Dogethereum construction retrofits Dogecoin and Ethereum without instituting a fork, change in mining protocol, or delegating authority to external consensus.  Dogecoin (resp.\ Ethereum) miners need not witness Ethereum (resp.\ Dogecoin) events.

\item \emph{Currency invariant.} The protocol is resilient against exchange rate fluctuations without appealing a price oracle.  Our assumption of rational actors guarantee that the number of actively circulating \WOW closely resembles the number of current ``locked'' \DOGE.  Moreover, each Operator's collateral remains constant, regardless of the exchange rate between \ETH and \DOGE.

\item \emph{Efficient.} The relay offers fast confirmation times at a low cost, both in terms of Ethereum gas and amount of Relayer deposits.  Crossers can obtain \WOW in minutes, and the protocol is snoozable in the sense that it only consumes gas when in use.

\item \emph{Secure.} We present an explicit and relatively simple construction (Section~\ref{sec:modules}), modulo bulletproof and Truebit black boxes, which is amenable to formal security analysis (Section~\ref{sec:illustrations}).  As we indicated earlier in this section, for example, the relay attacks resist orphaned Dogecoin block submissions.  Our two-way peg avoids distinguished and altruistic nodes (including the Bridge Contract creator), is robust against Sybil attacks and collusion, and relies exclusively on economic incentives for security.

\item \emph{Scalable.} Our two-way peg achieves cryptoeconomic security with any non-negative number of participants, however a greater number of participants means even more entwined eyes on the system.  The number of existing \WOW tokens does not limit transaction throughput, and the protocol can handle many simultaneous crossings.

\item \emph{Robust.} Dogethereum features an open network for anyone to immediately participate as Operator, Crosser, Hodler, Relayer, or Reporter with minimal capital and compute resources.  The construction is based on unanimous consensus and offers fair market operation to all participants.  Operators' collateral eventually gets returned, so long as the two-way peg remains in use.  During periods of moderate price stability, \WOW tokens are low-maintenance and behave exactly like standard ERC-20 tokens.
\end{enumerate}

\section{Modules specification} \label{sec:modules}

\begin{figure}[h!]
\begin{tabular}[b]{ll}
\hspace{-4ex}
\begin{tikzpicture}
\matrix (m) [matrix of nodes, row sep=2em, column sep=1.5em, font={\small \tt}]
{
\ & \boxed{\text{\hyperref[sec:onetime]{Genesis}}}\\
\boxed{\text{\hyperref[sec:bridgeopening]{Bridge opening}}}  &  & \\
\boxed{\hyperref[sec:locking]{\text{Locking}}}  & \hspace{-11ex} \qquad \boxed{
\begin{tabular}{c}
\textrm{\small Backstop modules:}\\[1ex]
\boxed{\text{\hyperref[sec:reportingmissing]{Reporting missing \DOGE}}}\\[1.2ex]
\boxed{\text{\hyperref[sec:backtracking]{Backtracking}}}
\end{tabular}
} &\  \\
\  & \ &\boxed{\text{\hyperref[sec:relay-listening]{Relay (Listening)}}}  \\
\  & \ &\boxed{\text{\hyperref[sec:relay-verification]{Relay (Verification)}}}  \\
\  &\boxed{\text{\hyperref[sec:minting]{Minting}}}  \\
\  &\boxed{\text{\hyperref[sec:unlocking]{Unlocking}} } \\
};

\path
(m-1-2) edge [->,thick=3pt, bend right=10] (m-2-1)
(m-2-1) edge [->,thick=3pt] (m-3-1)
(m-3-1) edge [->,thick=3pt, bend right] (m-6-2)
(m-1-2) edge [->,thick=3pt, bend left=30] (m-4-3)
(m-4-3) edge [->,thick=3pt] (m-5-3)
(m-5-3) edge [->,thick=3pt, bend left=10] (m-6-2)
(m-6-2) edge [->,thick=3pt] (m-7-2)
(m-5-3) edge [->,thick=3pt, bend left] (m-7-2)
(m-5-3) edge [->,thick=3pt, bend left] (m-3-2)
(m-3-2) edge [->,thick=3pt, bend left=20] (m-4-3);
\end{tikzpicture}

&
\end{tabular}
\caption{Arrows indicate module sequences over the course of a \DOGE lifecycle.  The vertical axis represents time, modulo the center box. \label{fig:modules}}
\end{figure}
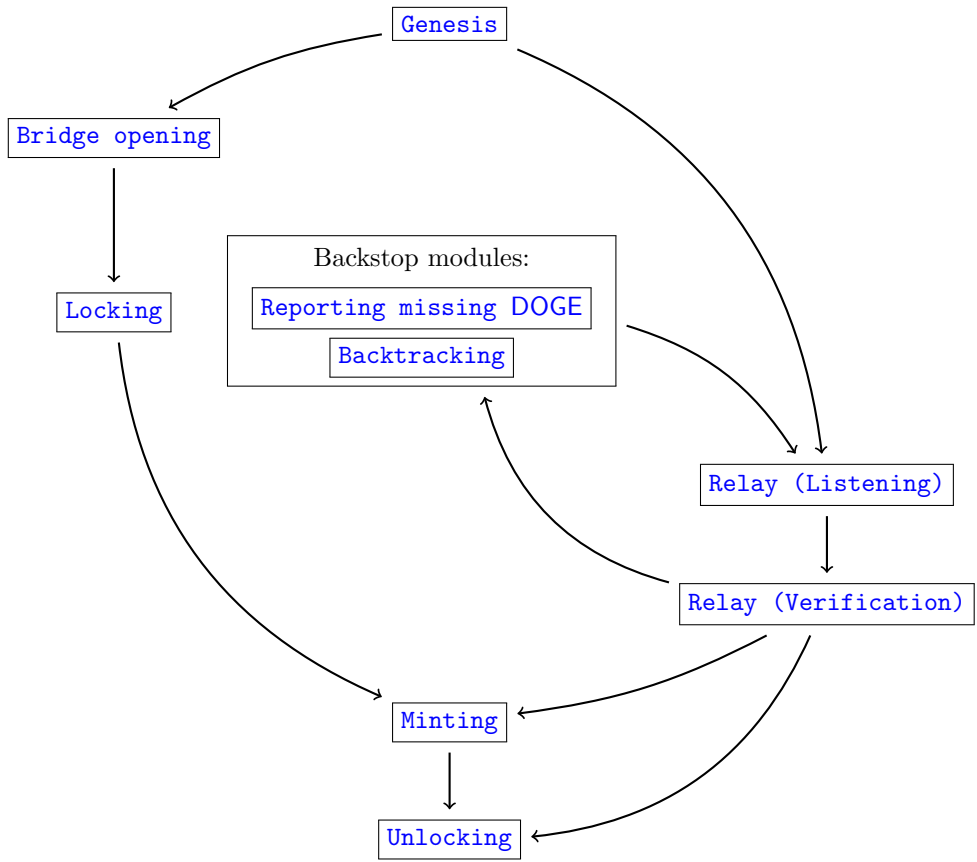

The two-way peg consists of nine modules.  We enumerate these in Figure~\ref{fig:modules}.  The major components include the collateralized peg (Bridge opening, Locking, Minting, \& Unlocking), it's relay (Listening \& Verification), and backstops (Reporting missing DOGE \& Backtracking).  As the relay involves timing coordination between Dogecoin and Ethereum, we first establish some relevant notation.

\begin{defn}
Let \emph{Ethereum time~$e$} denote the (approximate) clock time when the most recent Ethereum block has ordinal number~$e$.  We assume that network propagation is such that this quantity is well-defined (plus or minus 1 block).

Let us fix~$c$ and $d$ as universal, positive, integer-valued constants.  We shall say that a Dogecoin block is \emph{confirmed} when the network has propagated at least $c$~valid proof-of-works on top of it, and a Dogecoin block is \emph{recently confirmed} at Ethereum time~$e$ if its ordinal number plus $d$ exceeds the ordinal number of the most newly confirmed Dogecoin block at Ethereum time~$e$.  
\end{defn}

For simplicity of presentation, we shall assume that the Dogecoin block difficulty remains constant over time.  In reality, Dogecoin adjusts its proof-of-work difficulty after every block and targets one-minute intervals between blocks~\cite{dogecoin-github}.  More formally, in order to accommodate difficulty changes, the notion of maximal should be based on cumulative work rather than length, since, in reality, the ``true'' Dogecoin blockchain is not the longest one but the one with the greatest cumulative work.  Each Merklized sequence of proof-of-works should therefore include a measure of cumulative work and, in places where we write that the protocol compares lengths, it should really compare cumulative work.  Formally, where we say ``block extension of length~$c$,'' we really mean ``an extension of blocks whose cumulative proof-of-work is equivalent to the work of $c$ blocks according to Dogecoin's difficulty at the time of the first block in the sequence.''

The Bridge Contract either ignores or penalizes malformed or untimely interactions.  When such actions are clear from context, the protocol description may not state its course explicitly.

\subsection{Genesis}  \label{sec:onetime} 
 Before anything else happens, a universal \emph{Bridge Contract} opens in Ethereum for managing \ETH deposits, recording and coordinating verification for ``locks'' in Dogecoin, and minting \WOW tokens.  
\begin{itemize}
\item The Bridge Contract's \emph{history}, or record of Dogecoin events, begins in an empty state and a \emph{current date}, or latest known Dogecoin block, of~0.  

\item The \emph{Relay module} toggle initializes in ``Listening'' mode, see  ``Relay (Listening)'' below.

\item For each rational number~$y$, the first-in-first-out \emph{$y$-bridge queue} initializes in an empty state.
\end{itemize}
A relay initializing today would be behind on years' worth of Dogecoin blocks, however there is no need to catch up on their transactions because no ``locks'' could have taken place prior to this initialization.  Dogethereum need not know who owns which \DOGE at its inception nor how these these coins were obtained, and therefore the two-way peg does not track this information.  Barring any unfounded objections from onlookers (Section~\ref{sec:relay-verification}), the creator of the Bridge Contract may simply provide a first ``valid'' extension of blocks to the relay saying that no Dogecoin transactions have yet occurred and matching the Bridge Contract's current date with the ``true'' Dogecoin blockchain.  As we shall see, this process is uniformly consistent with the backtracking process outlined in Section~\ref{sec:backtracking}.

\subsection{Bridge opening} \label{sec:bridgeopening}
Any \emph{Operator} may establish a channel for converting \DOGE into \WOW by posting \ETH collateral into the Bridge Contract.  More formally, a \emph{bridge} is an \ETH collateral deposit of some amount~$x$, sent from the Operator's \ETH address, plus a quintuple whose components correspond to the following semantics:
\begin{enumerate}
\item a value~$y$ indicating the maximum exchange rate of the form $1~\ETH = y~\DOGE$ for which the deposit~$x$ suffices to collateralize against a Crosser's ``locked'' \DOGE,
\item a \DOGE address to hold the ``lock'' (controlled by the Operator),
\item the Operator's \emph{crossing fee}, measured in $\WOW[y]$, for transforming a Crosser's \DOGE into $\WOW[y]$,
\item (optionally) specifying a minimum number of \DOGE to initiate a ``lock,'' and

\item (optionally) a \emph{burn reporting bounty}, where the latter is a reward paid to anyone who later evidences to the Bridge Contract that the Operator appropriately released the ``locked'' \DOGE.
\end{enumerate}
The \emph{capacity} of this bridge is $x / y$, and the \emph{head} of the bridge is the Dogecoin address in Item~2.  An Operator \emph{opens} a bridge by sending collateral of $x$~\ETH and the data above to the Bridge Contract.

\subsection{Locking}  \label{sec:locking}
A \emph{Crosser} who wishes to convert \DOGE into $\WOW[y]$ ``\emph{locks},'' \DOGE by sending these funds to any open bridge head whose associated exchange rate~$y$ exceeds the ``true'' current exchange rate by a comfortable margin (as determined by the Crosser). 

Prior to ``locking'' \DOGE, the Crosser may \emph{register} with the Bridge Contract his intention to use a particular bridge head.  Registration ensures that each bridge head gets used at most once and avoids the situation where two Crossers submit \DOGE at the same time and a rational Operator walks off with the excess.  Once the Bridge Contract confirms registration, along with a small deposit, the Crosser has, say, 20 Dogecoin blocks within which to ``lock'' his \DOGE as measured by the history at confirmation time.  If the Crosser does not complete the transaction within this time interval, the registration is canceled and the Crosser pays from his deposit an \ETH \emph{registration voiding fee}, equivalent to a small percentage of the \DOGE he intended to ``lock'' at exchange rate~$y$.  This fee simply aims to discourage denial-of-service; honest Crossers receive a full refund of their deposit, less gas expenses.  Formally, registration is optional; in case of simultaneous crossings at a single bridge head, the Bridge Contract simply mints \WOW according to the first report that it hears about (Section~\ref{sec:minting}).

By sending \DOGE to the open bridge head, the Crosser commits to pay the following fees in $\WOW[y]$:
\begin{itemize}
\item the bridge's crossing fee, which may be a function of the number of \DOGE sent by the Crosser,

\item a fixed \emph{relay tax}, paid by the Crosser to the Relayer who successfully communicates the Dogecoin block containing the ``lock'' to the Bridge Contract (Section~\ref{sec:minting}), and

\item an optional \emph{lock reporting bounty} offered to whoever provides a Merkle proof that the Crosser's ``lock'' occurred (Section~\ref{sec:minting}).
\end{itemize}
We shall assume that the Crosser always sends a quantity of \DOGE equal to the bridge capacity.  If the  Crosser sends excess \DOGE, then the Crosser keeps the surplus.  If the Crosser sends too few \DOGE, then the Bridge Contract refunds some of the Operator's \ETH deposit.

\subsection{Relay (Listening)}  \label{sec:relay-listening}
We describe the process by which the system conveys events on Dogecoin to the Bridge Contract.  \emph{Relayers} convey updates from Dogecoin to the Bridge Contract's history while the Relay module toggles between two mutually exclusive modes called ``Listening'' and ``Verification.''  The present Listening module ends with a call to the Verification module described below.

\begin{defn}
A \emph{valid extension} of the Bridge Contract's history at Ethereum time~$e$ consists of a triple comprised of the following elements which aim to satisfy, respectively, the properties of Maximality, Validity, and Shallow-fork-free from Section~\ref{sec:bridgeoperation}.  The parameter~$d$ introduced in the beginning of Section~\ref{sec:modules} formalizes the notion in Item~1 below.
\begin{enumerate}
\item The ordinal number~$b$ of a recently confirmed Dogecoin block at Ethereum time~$e$.  We say that $b$ is the \emph{range} of the valid extension.

\item The Merkle hash of the Dogecoin blocks from the history's current date at Ethereum time~$e$ up through block number~$b$. We call this root the \emph{commitment} of the valid extension.

\item The Merkle hash of $c$~blocks witnessing that the commitment in Item~2  has been confirmed in Dogecoin.  We call this root the \emph{confirmation witness}.
\end{enumerate}
Item~3 is only used in the Verification module in case of a challenge.
\end{defn}
Any party that submits sufficient deposit to the Bridge Contract becomes a \emph{Relayer} and earns the right to append valid extensions to the Bridge Contract's history.  At any time, a Relayer may withdraw her deposit at the cost of losing Relayer status.  The required deposit amount to become a Relayer equals the cost of issuing a Truebit task to check a bulletproof witnessing Items 2 \& 3 above (see Sections~\ref{sec:relay-verification} and~\ref{sec:deposits} for more details). 
Note that Operator deposits used for bridge opening do not count towards Relayer deposits.  

The first Relayer who submits an alleged valid extension, or \emph{submission}, becomes the \emph{active} Relayer and causes the the Relay module to switch into ``Verification'' mode (Section~\ref{sec:relay-verification}).  Until then, the Relay remains in ``Listening'' mode.  We remarked earlier that Crossers, Hodlers, and Operators have incentive to participate as Relayers (Section~\ref{sec:relayers}) even without taking into consideration the relay tax reward.

\subsection{Relay (Verification)} \label{sec:relay-verification}

The Verification module provides the opportunity for any Relayer to challenge the active Relayer's submission. Let $k<d$ be a a small but fixed security parameter.  If no challenges occurs within $d-k$ Dogecoin blocks, as measured in (expected) equivalent Ethereum time, then Verification times out and the system \emph{accepts} the submission.  The point of the parameter~$k$ is to prevent a malicious Relayer from  knowing a longer chain in the future and using that to prove that an earlier submission wasn't maximal (Section~\ref{sec:relayattacks}).  In case of acceptance, the Bridge Contract:
\begin{enumerate}
\item appends the submission's commitment to the history and
\item updates the current date to the value from the submission's range.
\end{enumerate}

Let us now consider the case where some Relayer disagrees with the submission which, for the remainder of this subsection, we assume occurred at Ethereum time~$e$.  Such a Relayer is called a \emph{Challenger}.  Formally, a \emph{challenge} consists of a indication as to whether:
\begin{itemize} 
\item the submission's range is too small, in which case the challenge also includes an (alleged) new valid extension of the Bridge Contract's history at the current Ethereum time, or
\item  the submission's commitment is faulty.
\end{itemize}
Each challenge chooses exactly one of these options.  A claim of the first type indicates that none of the blocks in the submission were recently confirmed, while a claim of the second type means that the Merkle hash of the alleged Dogecoin blocks does not represent a valid proof-of work chain.  As a commitment might include orphaned blocks, a Challenger should always prefer a challenge of the first type.  Indeed an unrecognized Merkle hash with short range need not constitute an errant commitment.
 
The Bridge Contract only considers the first challenge that it receives, unless the challenge fails in Step~1(a) below, in which case it continues to wait for further challenges.  There are two cases.
\begin{enumerate}
\item Assume that the Challenger claims that the range is too short.\begin{enumerate}
\item If the range of the Challenger's (alleged) valid extension exceeds the submission's range by less than~$d$,  then the challenge is simply ignored and discarded.
\item Otherwise, the (alleged) valid extension provided by the Challenger:
\begin{enumerate}
\item replaces the submission,
\item the Challenger becomes the active Relayer,
\item the Verification module restarts, and 
\item the Relayer pays some small penalty to discourage denial-of-service but has the option to contest the validity of the Challenger's new submission, via Truebit (see below), in exchange for a refund but at the potential cost of losing her remaining deposit.
\end{enumerate}
\end{enumerate}

\item Now assume that the Challenger claims that the commitment is faulty.  Then the active Relayer must provide a proof of its correctness, namely a bulletproof witnessing the following properties of the submission.
\begin{itemize}
\item The sequence of blocks represented in the commitment form a valid sequence of Dogecoin proof-of-work of length prescribed by the range.

\item The first block represented in the commitment is a valid extension of the latest block in the Bridge Contract's history at  Ethereum time~$e$.  (If the history is empty, as it would be at genesis, ignore this step).

\item The sequence of blocks represented in the submission's confirmation witness: 
\begin{itemize}
\item is a valid sequence of Dogecoin proof-of-work of length~$c$ and
\item extends the commitment.
\end{itemize}
\end{itemize}
\end{enumerate}

While the active Relayer prepares the bulletproof, which could take a while (Section~\ref{sec:bulletproofs}), the Bridge Contract splits into two threads.  One thread returns the relay to a Listening state, without adding the submission to the history, while the other submits the bulletproof, upon receipt, to Truebit for verification. (If the bulletproof is not forthcoming within the timeout period, then the Bridge Contract destroys the active Relayer's deposit.)

The remainder of the thread determines whether the Challenger or the active Relayer pays for the Truebit task,  Either way, the Bridge Contract does not add the submission to its history.  There are two disjoint possibilities.

\begin{enumerate}
\item If Truebit accepts the bulletproof, then the Bridge Contract uses the Challenger's deposit to pay for the Truebit task and to compensate the vindicated active Relayer.
\item If Truebit rejects the bulletproof, then the Bridge Contract uses the active Relayer's deposit to pay for the Truebit task and to reward the successful Challenger. 
\end{enumerate}
The Verification module thread terminates.

\subsection{Minting} \label{sec:minting}

\emph{Reporters} reveal ``locks'' to the Bridge Contract by unpacking commitments  stored in its history.  Reporters need not post deposits.  Formally, a \emph{report} consists of a canonical representation of the following:
\begin{itemize}
\item a pointer to a commitment in the history, and
\item a Merkle proof for a Dogecoin transaction contained within the commitment.
\end{itemize}
Each Dogecoin transaction is ``unused'' by default.  The minting module proceeds as follows.
\begin{enumerate}
\item A Reporter sends a lock report to the Bridge Contract.  The Bridge Contract then checks whether the report meets the following conditions.
\begin{enumerate}
\item The indicated commitment exists and the the Merkle proof points to a valid transaction.
\item The receiver of the transaction is a bridge head whose bridge corresponds to exchange rate~$y$ (for some~$y$).
\item The sender of the transaction is registered as a Crosser for this bridge head, if any registration has taken place.
\item The transaction is ``unused.''
\end{enumerate}
\item If all conditions are met, then the Bridge Contract:
\begin{enumerate}
\item mints to the Crosser's Ethereum address a quantity of $\WOW[y]$ equal to the bridge's capacity,
\item adds the bridge to the end of the $y$-bridge queue, 
\item pays:
\begin{enumerate}
\item the lock reporting bounty to the Reporter and
\item the relay tax to the Relayer who committed the encompassing block to the history, and
\end{enumerate}
\item permanently marks the transaction as ``used.''
 \end{enumerate}
 \item Otherwise, the Bridge Contract ignores the lock report.
\end{enumerate}

\subsection{Unlocking} \label{sec:unlocking}

When a \emph{Hodler} \emph{burns} $\WOW[y]$ into the Bridge Contract, thereby destroying the tokens, the Operator whose bridge lies at the front on the $y$-bridge queue becomes obligated to send an equal number of \DOGE to a Dogecoin address specified by the Hodler.  For simplicity, the Operator pays this Dogecoin transaction fee.  

A Hodler may either claim more or less than the \DOGE balance stored at the Operator's bridge head.  At the time of a bridge's creation, its \ETH collateral equals its capacity divided by~$y$.  The bridge's \ETH collateral is now decreased by the amount of $\WOW[y]$ burned by the Hodler divided by~$y$ and placed into \emph{escrow}.  In Section~\ref{sec:reserve} we shall argue that the the $y$-bridge queue maintains sufficient collateral to cover the Hodler's exchange request.

In case the Hodler's $\WOW[y]$ burn amount exceeds the number of \DOGE ``locked'' in the head of the $y$-bridge queue's first element, then the Hodler's \DOGE request is distributed over successor bridges as well.
\begin{enumerate}
\item \ETH collateral is recursively distributed into escrow among successive bridges in the $y$-bridge queue, and
\item each bridge whose head's balance reaches~0 is popped from the $y$-bridge queue.
\end{enumerate}

Upon the Hodler's burn, the Bridge Contract  simultaneously does the following for each bridge involved in the above recursion.
\begin{enumerate}
\item It waits until some designated timeout for a report witnessing the Operator's \DOGE payment to the Crosser.  The Bridge Contract does the following for each report.
\begin{enumerate}
\item It checks whether submitted report meets all of the following criteria.
\begin{enumerate}
\item The indicated commitment exists and the Merkle proof points to a valid transaction.
\item The sender of the transaction is the Operator and the receiver is the address indicated by the Hodler.
\item The Dogecoin transaction was reported after the Holder burned $\WOW[y]$.
\end{enumerate}
\item If these criteria are met,
\begin{enumerate}
\item the Reporter receives the burn report bounty(s),
\item the Operator receives a refund of her \ETH deposit, less the remaining balance, and
\item the Unlocking module ends.
\end{enumerate}
\item Otherwise the report is ignored.
\end{enumerate}
\item If no successful report occurs before the timeout, the Bridge Contract pays the corresponding fraction of the Operator's \ETH collateral to the Hodler.
\end{enumerate}

\subsection{Reporting missing \DOGE} \label{sec:reportingmissing}
For purposes of security, the Bridge Contract includes two backstop modules.  These modules never execute unless some party deviates from anticipated rational behavior.  At any moment, an Operator can walk off with ``locked'' \DOGE, and any \WOW Hodler can then burn \WOW in exchange for his \ETH collateral.  The formal process is as follows.
\begin{enumerate}
\item The Hodler submits a report to the Bridge Contract demonstrating that the Operator moved ``locked'' \DOGE out of a bridge with exchange rate~$y$.  For purposes of detection, we assume that the Operator uses the bridge head exclusively for Dogethereum.

\item The Hodler burns~$n$ $\WOW[y]$ tokens into the Bridge Contract.

\item If both:
\begin{enumerate}
\item the report is valid and shows that the Operator moved $n$~\DOGE, and
\item  the report's transaction is ``unused,''
\end{enumerate}
\item then the Bridge Contract
\begin{enumerate}
\item sends $n/y$ \ETH from the Operator's deposit to the Hodler,
\item reduces the bridge's \ETH collateral by the same amount, and
\item marks the report's transaction as ``used.''
\end{enumerate}
\item Otherwise, the report is ignored.
\end{enumerate}

\subsection{Backtracking} \label{sec:backtracking}
If the Bridge Contract for some reason appends a commitment to the Bridge Contract's history which does not reflect the true Dogecoin blockchain, no honest Relayer can further extend the history using the protocol above.  For this reason, the Bridge Contract permits Relayers to extend its history from any previous history state.  For purposes of consistency, however, only ``locks'' and ``unlocks'' in blocks with higher ordinal number than the Bridge Contract's current date will impact present and future Bridge Contract actions.  Backtracking might become necessary, for instance, in a moment where no Operators, Crossers, and Hodlers are watching and an attacker cheaply halts operation through a bogus history extension.  We remark, however, that an attacker does not gain coins or tokens via this method when no \DOGE are locked and no \ETH collateral is staked, which is the most likely scenario where no agents are watching.

The size of the Relayers' deposits (Section~\ref{sec:deposits}) bounds the depth of secure backtracking.  This finite bound may suffice for short-term corrections, but in cases where the relay remained inactive for a longer period of time, or in its its moment of genesis (Section~\ref{sec:onetime}), a deeper backtracking or accelerated extension method may be necessary.  Unfortunately, in such cases, the system lacks the resources to determine whether or not a challenge to a very long extension is justified, and thus any objection must suffice to break unanimous consensus without further scrutiny.  Thus the relay may include two deeper backtracking modes beyond the simple one described in the previous paragraph as follows.
\begin{enumerate}
\item Anyone may propose an extension or backtracking of any length, however any objection or normal extension within, say, a 24-hour period suffices to cancel that extension.

\item In case 72 hours go by without any progress on Item~1, the two-way peg may permit an even deeper backup mode which allows extensions in chunks which are short enough to fall back on standard deposits for security.
\end{enumerate}

Reinitializing the bridge from a deeply stuck state essentially requires unanimous consensus, and the method above could result in a permanently stuck state in case an attacker successfully places a long, bogus extension with some range that is years into the future.  Presumably, however, this permanent stuck state happened because no one was actually using the bridge, hence it is not a disaster.

\begin{figure}
\begin{tabular}{p{6.2cm}|p{4.6cm}}
\textbf{parameter} & \textbf{suggested value} \\
\hline
-confirmation depth $c$ & $\approx 10$ Dogecoin blocks \\
-recently confirmed threshold $d$ & $\approx 20$ Dogecoin blocks\\
-security parameter $k$ & $\approx 2$ Dogecoin blocks\\
-Relayer deposit & see Section~\ref{sec:deposits} \\
-registration void & $\approx 1\%$ of cross in \ETH\\
-time to ``lock'' after registration & $\approx$ 20 Dogecoin blocks \\
-penalty for non-maximal extension & $\approx$ 10\% of \ETH deposit\\
-timeout to pay ``unlock'' & $\approx 20$ Ethereum blocks\\
-timeout for challenge & linear in $d$ \\
-bulletproof construction timeout & linear in length of extension\\
-maximum extension length & $\approx$ 10,0000 Dogecoin blocks \\
\begin{tabular}{l}
-reward for correct challenge/ \\
compensation for vindicated Relayer
\end{tabular}
  & $\approx 1\%$ of verification cost\\
-deep backtracking delays & 24 \& 72 hours\\
\end{tabular}
\caption{Protocol parameters and suggested values. \label{fig:parameters}}
\end{figure}

\section{Illustrations} \label{sec:illustrations}

We explore a few scenarios and additional features which explain the workings of the two-way peg.

\subsection{Reserve equilibrium} \label{sec:reserve}

Let us quantify the extent to which Dogethereum's cryptoeconomic incentives protect the interests of rational agents.  We shall argue that agents who faithfully monitor the \ETH to \DOGE exchange rate and events on Dogecoin blockchain, according to the standards described in Section~\ref{sec:coc}, never lose value due to fractional reserves among \DOGE, \ETH, or \WOW.

Let us first observe that the Bridge Contract alone directly anchors the reserve ratio between \ETH and \WOW.  As the Bridge Contract has sole authority to both mint and burn \WOW tokens and uniquely escrows all \ETH collateral, it necessarily monitors all bookkeeping activities for these two currencies.  The Bridge Contract only mints $\WOW[y]$ tokens (Section~\ref{sec:minting}) after an Operator drops \ETH collateral into it at exchange rate~$y$ (Section~\ref{sec:bridgeopening}), and it only releases \ETH collateral when an equivalent number of $\WOW[y]$ tokens get burned at exchange rate~$y$, either through Unlocking (Section~\ref{sec:unlocking}) or Reporting missing \DOGE (Section~\ref{sec:reportingmissing}).  Thus we have observed the following.
\begin{invariant} \label{inv:wow-eth}
The number of $\WOW[y]$ in circulation equals $y$ times the number of \ETH collateralized at exchange rate~$y$.
\end{invariant}
Both here and below, the phrase ``in circulation''  orients the reader from the perspective of the Bridge Contract.  Invariant~\ref{inv:wow-eth} does not take into account Hodlers who either stuffed $\WOW[y]$ under a mattress or accidentally lost their private keys.  

We now ascertain agents' benefits from Invariant~\ref{inv:wow-eth}.  First, let us understand why any Hodler who faithfully tracks exchange rates and burns $\WOW[y]$ in a timely manner can recover a fair amount of \ETH in case ``locked'' \DOGE has become unavailable.  By virtue of the fact that the Hodler possesses $\WOW[y]$, it follow from Invariant~\ref{inv:wow-eth} that the Bridge Contract holds some Operator's \ETH collateral of equal or greater value.  Now consider the particular such Operator(s) whose bridge(s) lie at the front of the $y$-bridge queue.  In case the Hodler attempts to ``unlock'' \DOGE which is not forthcoming from the Operator(s), then the Bridge Contract ensures that the Operator(s)' \ETH collateral is available for the Hodler's collection (Section~\ref{sec:unlocking}).  Thus we obtain the the following.
\begin{invariant} \label{inv:hodler}
A Hodler who believes that the value of 1~\ETH exceeds the value of $y$~$\WOW[y]$ never loses net value by exchanging $\WOW[y]$ through the Bridge Contract.  In more detail, A Hodler can, at any moment, burn a quantity of, say, $w > 0$ $\WOW[y]$ in exchange for some amount~$0 \leq d \leq w$ of ``locked'' \DOGE plus $(w-d)/y$ collateralized \ETH.  The Hodler may choose any value $w$ within his means, however the parameter $d$ depends on the current state of the system.
\end{invariant}

By inspection, Operators, Crossers can preserve value as well by monitoring Dogecoin and challenge as Relayers in case of bogus extensions.  Since the Operator's \ETH collateral is directly recorded in the Bridge Contract (Section~\ref{sec:bridgeopening}), the Bridge Contract
can correctly manage the release of this \ETH collateral upon properly learning of an ``unlock'' assuming the Operator's ``locked'' \DOGE did not go missing (Section~\ref{sec:unlocking}).  Similarly, a Crosser receives \WOW immediately upon reporting of a ``lock'' (Section~\ref{sec:locking}).  Deposits of honest Relayer deposits are protected via the correctness of bulletproofs and Truebit operations, as spending of Relayer deposits occurs exclusively as a result of bulletproof verification (Section~\ref{sec:relay-verification}).  Moreover Relayers who neither challenge nor broadcast valid extensions cannot lose their deposits.

As a sanity check, let us review Invariant~\ref{inv:hodler} in the context of rational actors with an eye towards the desired bijection between ``locked'' \DOGE and \WOW tokens.  Consider the two possible cases where an Operator walked off with ``locked'' \DOGE.  Without loss of generality, assume that the \DOGE were ``locked'' into a bridge with exchange rate~$y$.
\begin{quote}
\textit{\textmd{Case 1:} }The Operator's \ETH collateral exceeds the ``true'' value of the formerly ``locked'' \DOGE.
\end{quote}
In this case, a rational Hodler will gladly burn $\WOW[y]$ in exchange for the Operator's \ETH collateral, which exists by Invariant~\ref{inv:wow-eth}.  Then there's no \DOGE deficit incurred.
\begin{quote}
\textit{\textmd{Case 2:}}  The value of the Operator's formerly ``locked'' \DOGE exceeds the Operator's \ETH collateral.
\end{quote}
In this case, the $\WOW[y]$ bridge collapses in the sense that the protocol doesn't care what happens to $\WOW[y]$ tokens anymore.  Everyone has switched to using $\WOW[y/2]$ tokens at this point, and any remaining \linebreak $\WOW[y/2]$ Hodlers inherit the consequences of not cashing out to \DOGE soon enough.  If in the future the price of \ETH rises again, then we are back again in Case 1 above, which then returns the $\WOW[y]$ token supply to equilibrium.

In summary, if we add rationality assumptions in Invariant~\ref{inv:hodler}, then we effectively obtain the tighter bound $d=w$ since any \ETH collateral would already have been obtained by a Hodler.
\begin{invariant} \label{inv:doge-wow}
Assume that the``true'' exchange rate , according to all Hodlers, is such that $1\ \ETH$ is worth at least $y$ \DOGE.  Further assume the following.
\begin{enumerate}
\item There exist rational Hodlers who report missing \DOGE (Section~\ref{sec:reportingmissing}).

\item Operators are rational in that they challenge Relayers, presumably Crossers, who attempt to relay bogus ``locks'' into the Bridge Contract (Sections~\ref{sec:locking} and~\ref{sec:relay-verification}).

\item Crossers are rational insofar as they collect their $\WOW[y]$ (Section~\ref{sec:minting}) after crossing (Sections~\ref{sec:locking}) and challenge Operators who might attempt to cheat them through bogus relay submissions.
\end{enumerate}
Then for every $y$, the number of ``locked'' \DOGE with exchange rate~$y$ on the ``true''  Dogecoin blockchain equals the number of $\WOW[y]$ in circulation.
\end{invariant}

It follows by transitivity from Invariants~\ref{inv:wow-eth} and~\ref{inv:doge-wow} that an equilibrium also exists between the number of ``locked'' \DOGE and the amount of \ETH collateral under the join of the respective assumptions.  We remark that Item~2 in Invariant~\ref{inv:doge-wow} is needed to mitigate against the unfortunate situation where an Operator places an \ETH collateral in the Bridge Contract and then a Crosser misuses the relay by lying about depositing funds into it.  In practice, this situation creates a \DOGE deficit until that bridge reaches the front of its respective bridge queue because there is no way to report the \DOGE as missing.

\subsection{Using \WOW in smart contracts} \label{sec:dex}

Unlike human Hodlers, who can detect changes in price volatility based on off-chain signals, smart contracts react exclusively to data and events on Ethereum.  Moreover, a smart contract remains dormant until a confirmed transaction incites a change.  Thus a smart contract could miss a critical change in the \DOGE--\ETH exchange rate and lose value in its $\WOW[y]$ holding.

In order to mitigate this problem, the smart contract using $\WOW[y]$ may provide an incentive for an outside party to wake it up with new information about exchange rates, at which time the smart contract might exchange its $\WOW[y]$ for some (possibly smaller quantity of) $\WOW[y/2]$.  Operators, who set the market rate for crossing their bridges (Section~\ref{sec:bridgeopening}), have incentive to make bridges with exchange rate $\WOW[y/2]$ when the time comes.

Ideally, the smart contract would confirm the new price via a decentralized exchange reflecting ``true'' market rates and without risking price manipulation through a set of centrally curated data feeds.  Decentralized exchanges remain a critical, open area for further research.

\subsection{Fine-tuning relay deposits} \label{sec:deposits}

Relay deposits must be sufficient high to pay for the cost of preparing an scrypt bulletproof and verifying it with Truebit (Section~\ref{sec:relay-verification}), but how much is this exactly?  The actual cost could depend on many factors independent of the price of \ETH or Ethereum events, including the local cost of electricity needed to perform the task. Secondarily, the deposit must  also be large enough to cover the penalty for non-maximal extensions and rewards for correct challenges (Section~\ref{sec:relay-verification}).  We remark that bulletproofs reduce the deposit requirement by minimizing on-chain storage and verification resources.  

As the cost of verifying an extension scales with the number of proof-of-works it contains, a variation in the price of \ETH relative to the cost of electricity could, assuming a fixed amount for the Relayer's deposit, affect the maximum extension length.  One might prefer to have a fixed deposit, rather than a variable maximum extension length based on each Relayer's chosen deposit amount, as the latter configuration could bias the system towards Relayers with the deepest pockets, particularly in cases of infrequent relay use.

Rather than trying to compose a comprehensive and fixed list of external factors which could potentially affect the minimal Relay deposit amount, we instead include this value, along with the maximum length of an extension, as a parameter in the \WOW token itself, similar to exchange rates (Sections~\ref{sec:nutshell} and~\ref{sec:reserve}).  Individuals and smart contracts can then independently decide how to interpret these parameters.  A simple heuristic, on which which users could even agree off-chain, might be to target a deposit amount equal to the \ETH equivalent of, say, the maximum of 5~USD or the cost to verify a 10,000-block extension.  Any \WOW whose deposit parameters reflect out-of-range deposit parameters might be abandoned, but Hodlers can exchange them for \DOGE before this happens.  We remark that changes in the deposit parameters might necessitate, for security reasons, a new relay to be initialized via Genesis (Section~\ref{sec:onetime}).

Other constants from Figure~\ref{fig:parameters} could be parameterized as well, although for purposes of simplicity or for new feature upgrades, one might simply choose to deploy a new Bridge Contract instead at the additional cost of deploying a fresh smart contract.

\subsection{Relay security} \label{sec:relayattacks}
While we explicitly designed the relay to mitigate against certain vulnerabilities (Section~\ref{sec:bridgeoperation}), its implementation (Sections~\ref{sec:relay-listening} and~\ref{sec:relay-verification}) merits further analysis.  The components of a \emph{valid extension} (Section~\ref{sec:relay-listening}) satisfy the Requirements  of Maximality, Validity, and Shallow-fork-free (Section~\ref{sec:bridgeoperation}) in a bijective fashion.  Hence the Bridge Contract's can safely append to its history a submission with these properties.  Any active Relayer that submits an extension to the Bridge Contract which breaks one of these three properties should expect to get challenged and have his submission rejected.  Let us now inspect some more subtle attack surfaces.

\paragraph{Maximality gap parameter.}
An active Relayer must provide a valid extension whose range is within~$d$ Dogecoin blocks of maximal (Section~\ref{sec:relay-listening}), however the Challenger must respond to any error within Dogecoin $d-k$ blocks, as measured in Ethereum time, for some security parameter~$k \approx 2$ (Section~\ref{sec:relay-verification}).  If the Challenger were allowed to wait longer, he might be able to successfully challenge an honest Relayer by providing a longer valid extension which was not available to the original Relayer at the time of her submission.  A higher value for~$k$ would defend against an adversary with more mining power who, by chance, gets ahead of the main chain via selfish mining~\cite{ES18,TJS17} as well as the ``Challenging as denial-of-service'' attack below.

\paragraph{High range extension.}
Suppose that an active Relayer provides a range far greater than what currently exists on the ``true'' Dogecoin blockchain.  The Relayer would then be unable to provide a correct Merkle hash for the commitment, since his claimed blocks have not yet been created.  Another Relayer could then recognize the mistake and challenge, and the active Relayer would lose his deposit for being unable to provide a correct bulletproof.  Hence this attack is not rational.

\paragraph{Challenging as denial-of-service.}
A small denial-of-service penalty applies when an active Relayer is challenged after providing a non-maximal extension, unless the Challenger was at fault (Section~\ref{sec:relay-verification}).  Repeatedly claiming non-maximality thus becomes costly.  There might be some risk that a persistent Challenger could repeatedly challenge himself until the relay reaches its maximum permitted extension length (Section~\ref{sec:deposits}), however, the attack would require substantial investment and might be thwarted by other Relayers who also submit valid extensions in the meantime.

\section{Scrypt bulletproofs} \label{sec:bulletproofs}

Dispute resolution for extensions submitted by Relayers requires an efficient way to prove validity of chains of Dogecoin proof of work; in particular Relayers must be able to efficiently prove that they have the pre-images of many instances of the scrypt hash function using Dogecoin mining parameters. We have chosen to use bulletproofs for this purpose, an efficient argument system for arbitrary arithmetic circuits which has been shown to be competitive with SNARK proof systems in both computational effort and proof size required~\cite{Wahby:2017aa}. Critically for our purpose of verifying entire chains of proof of work, bulletproofs have also been shown to be highly efficient for batch verification. This is done by reducing the verification of $m$ proofs into verifying a single relation which scales well using an efficient multi-exponentiation, so that relatively little marginal computation time is added for each additional proof even while the proof size increases as $O(\log(m))$. 

Verification of the scrypt hash function provides especial need for an efficient proof system, as its memory-hardness involves several thousand calls to the xor-heavy stream cipher salsa8, in addition to requiring thousands of randomized memory accesses, both of which result in a circuit with high multiplicative complexity~\cite{tarsnap}.

We mitigate the above issue by building an arithmetic circuit for scrypt verification customized for the Dogecoin mining parameters; In particular we omit parts of the function not touched using these parameters, such as PRF chaining in PBKDF2 key derivation, in order to reduce circuit size and improve efficiency. The circuit was built using xjsnark, an optimized high-level framework used for encoding arithmetic circuits which produces an extended variant of the format used by Pinocchio~\cite{xjsnark,Parno:2013aa}. Extrapolating from current benchmarks using batch verification implemented over the secp256k1 curve, we estimate that 1, 10, and 100 Dogecoin proofs-of-work can be verified in 12, 20, and 80 minutes respectively~\cite{libsecp256k1}. Further more efficient implementations may reduce this time considerably; an implementation using Rust has recently obtained a 2x speedup using parallel formulas in the Edwards 25519 curve. Arithmetic circuits for general proofs have not yet been implemented but development is ongoing~\cite{dalek-crypto}.

We note that our two-way collateralized peg can also be extended to work with other blockchains utilizing Nakamoto consensus. In particular, we can create a two-way peg between Bitcoin and Ethereum using the same methodology. Using a construction identical to the scrypt arithmetic circuit above, we also leverage a circuit for SHA256 with 25,344 multiplication gates, a slight improvement over the previously implemented jsnark circuit~\cite{BBBPWM17}. Concrete benchmarks using the secp256k1 curve yield verification times for 1, 2, and 100 proofs of 582.4ms, 899.2ms, and 4333.8ms respectively. With only 47ms additional work per proof, it would take just under 8 minutes to verify 10,000 instances of Bitcoin proof of work.

As bulletproofs require circuits to be converted to a set of linear constraints, preprocessing is currently a bottleneck in the benchmarking process~\cite{BCCGP16}. We are currently working on implementing a more efficient pipeline to obtain more concrete estimates of circuits with high multiplicative complexity.

\paragraph{Acknowledgements.}
This paper is the collective result of many fruitful discussions.  The author thanks Ismael Bejarano and Oscar Guindzberg for pointing out the orphaned block attack by sharing a preliminary version of their paper~\cite{superblocks} (Section~\ref{sec:bridgeoperation}), Patrick McCorry and Jonathan Bootle for suggesting the use of bulletproofs, Sunny Aggarwal and Jim Posen for introducing the notion of a collateralized peg (Section~\ref{sec:nutshell}), Ahmed Kosba for help in navigating his xjsnark framework,  Alex van de Sande for suggesting a decentralized exchange which led to stability for \WOW in smart contracts (Section~\ref{sec:dex}), and Jeff Coleman for the bringing to our attention the synergy between two-way pegs and atomic swaps (Section~\ref{sec:relatedwork}).  We also thank Christian Reitwie{\ss}ner and Sina Habibian for useful conversations.

\bibliographystyle{plain}
\bibliography{dogethereum}

\end{document}